\documentclass[]{procpavia}
\usepackage{epsfig}
\usepackage{psfig}
\usepackage{graphicx}
\usepackage{graphics}
\usepackage{xspace}
\usepackage{amssymb}
\usepackage{latexsym}
\usepackage{natbib}
\usepackage{mathrsfs}
\newcommand{\inieq}{\begin{eqnarray}}            
\newcommand{\fineq}{\end{eqnarray}}            
\newcommand{\ove}{\overline}                    
\newcommand{\diff}{{\rm\,d}}                    
\newcommand{\bint}{\mskip .5mu \int \mskip-18mu} 
\newcommand{\half}{_{1/2}}
\newcommand{\thalf}{_{3/2}}

\def\s{\mbox{\boldmath $s$}}

\def\N{\mbox{\boldmath $N$}}
\def\L{\mbox{\boldmath $L$}}                                  
\def\r{\mbox{\boldmath $r$}}

\def\P{\mbox{\boldmath $P$}}
\def\q{\mbox{\boldmath $q$}}

\def\T{\mbox{\boldmath $T$}}

\def\N{\mbox{\boldmath $N$}}

\def\ss{\mbox{\boldmath $\sigma$}}

\def\r{\mbox{\boldmath $r$}}
\def\pf{\mbox{\boldmath $p^{ \prime}$}}
\def\pm{\mbox{\boldmath $p_{\mathrm m}$}}
\def\j{\mbox{$\boldmath {j}$}}
\def\ep{\mbox{$e^{\prime}$}}
\def\mcg{\mbox{$\mathcal{G}$}}
\def\mcv{\mbox{$\mathcal{V}$}}

\begin{document}
\pagestyle{prochead}
\title{Relativistic approach to $(e,e^{\prime}p)$ and $(e,e^{\prime})$ 
reactions} 
\author{Andrea Meucci} 
\author{Carlotta Giusti}
\author{Franco Davide Pacati }
\affiliation{Dipartimento di Fisica Nucleare e Teorica, 
Universit\`{a} di Pavia and \\
Istituto Nazionale di Fisica Nucleare, 
Sezione di Pavia, I-27100 Pavia, Italy}


\begin{abstract}
A relativistic distorted wave impulse approximation 
model for electron-induced one proton knockout reactions and a Green's function
approach to inclusive scattering are developed. 
Results for $\left(e,e^{\prime}p\right)$ and $\left(e,e^{\prime}\right)$ 
reactions are presented in various kinematical conditions and compared 
(when possible) with nonrelativistic calculations.  
\end{abstract}

\pacs{25.30.Fj: Inelastic electron scattering to continuum, 
25.20.Lj: Photoproduction reactions, 
24.10.Jv: Relativistic models, 
24.10.Eq: Coupled-channel and distorted-wave models,
24.10.Cn: Many-body theory}

\maketitle

\section{Introduction}
\label{sec.int}
A long series of high-precision experiments on several
nuclei~\cite{frou,mou,bern,dewitt,lapik,blomq,nikhef} have generated  
a well established tradition which singles out exclusive $(e,e'p)$ knockout
reactions as the primary tool to explore the single-particle aspects of the
nucleus.
Theoretical calculations were carried out within the  
framework of a 
nonrelativistic distorted wave impulse approximation (DWIA), where final-state 
interactions (FSI) and Coulomb distortion of the electron wave functions are 
taken into account \cite{DWEEPY}. Phenomenological ingredients were used to 
compute bound 
and scattering states. 
This approach was able to describe to a high degree 
of accuracy the shape of the experimental momentum distribution for
several nuclei in a wide range of different kinematics~\cite{book,kellyrep}.
However, a systematic rescaling of the normalization of the 
bound state, interpreted as the spectroscopic factor for the corresponding
level, had to be applied in order to reproduce the magnitude of the experimental
distribution.

Similar models based on a fully relativistic DWIA (RDWIA) framework were 
developed in more recent years. The Dirac equation is solved directly 
for the nucleon bound and scattering states~\cite{pickvanord,jinonl,spain}
or, equivalently, a Schr\"odinger-like equation is solved 
and the spinor distortion by the Dirac scalar and vector potentials is
incorporated in an effective 
current operator in the so-called effective Pauli 
reduction \cite{heda,kelly,mgp}. 

A successful description of new $(e,e'p)$ data at higher momentum transfer 
from Jefferson
Laboratory (JLab)~\cite{e89003,e89033} has been achieved within the framework of
RDWIA, but slightly different
spectroscopic factors are deduced, because the 
relativistic optical potentials in general give a stronger absorption than 
the corresponding nonrelativistic ones~\cite{spain,bgpc}.
Moreover, the limits of validity of the nonrelativistic 
DWIA analysis versus RDWIA were not always properly explored, as discussed in
Ref.~\cite{mgp}, resulting, e.g., in a certain degree of ambiguity for
the spectroscopic factors extracted at low energy.

In Sec. \ref{sec.eep}  
a RDWIA approach to low- and high-energy $\left(e,e^{\prime}p\right)$ data is 
presented and a careful
analysis of the limits of the nonrelativistic DWIA
is carried out. The sensitivity to different off-shell prescriptions
for the electromagnetic current operator will also be discussed~\cite{meu,mrd}.

In Sec. \ref{sec.ee} a 
Green's function approach for the inclusive electron scattering is
developed \cite{ee}. 
In the inclusive $(e,e^{\prime})$ scattering only the scattered electron is
detected whereas the final nuclear state is not determined. The
one-body mechanism is assumed to give the main contribution to the reaction in 
the quasielastic region. However, when the experimental data of the separation
between the longitudinal and transverse responses became available it was clear
that a more complicated framework than the single-particle model coupled to
one-nucleon knockout was necessary.
A review till 1995 of the experimental data and their possible explanations is
given in Ref. \cite{book}. Thereafter, only a few experimental papers were 
published \cite{batesca,csfrascati}. 
New experiments with high experimental resolution are planned at 
JLab \cite{jlabpro}  in order to extract the response functions.

From the theoretical side, a wide literature was produced in order to explain
the main problems raised by the separation, i.e., the lack of strength in 
the longitudinal response and the excess of strength in the transverse one.
The more recent papers are mainly concerned with the contribution to the
inclusive cross section of meson
exchange currents and isobar excitations \cite{Sluys,Cenni,Amaro}, with
the effect of correlations \cite{Fabrocini,Co}, and the use
of a relativistic framework in the calculations \cite{Amaro}.

At present, however, the experimental data are not yet completely understood. 
A possible solution
could be the combined effect of two-body currents and tensor correlations
\cite{Leidemann,Fabrocini,Sick}.

In our 
Green's function approach to $(e,e^{\prime})$ scattering the spectral 
representation of the
single particle Green function, based on a biorthogonal expansion in terms 
of the eigenfunctions of the nonhermitian optical potential, allows one to 
perform explicit calculations and to treat final state interactions 
consistently in the inclusive and in the exclusive reactions.

\section{The $\left(e,e^{\prime}p\right)$ reaction} \label{sec.eep}
\subsection{Relativistic current}
The main ingredient of the calculation is the nuclear transition 
amplitude, i.e.,  
\begin{eqnarray}
J^{\mu} = \int \diff \r \ove \chi^{(-)}(\r) \widehat j^{\mu} 
\exp{\{i\q\cdot \r\}}
 \varphi(\r) \ . \label{eq.rj}
 \end{eqnarray}
In RDWIA it is calculated using relativistic wave functions for initial and 
final states.

The bound state wave function, $\varphi$, is given by the Dirac-Hartree 
solution of a relativistic Lagrangian
containing scalar and vector potentials deduced in the context of a 
relativistic mean field theory that
satisfactorily reproduces single-particle properties of several spherical and
deformed nuclei \cite{lala}. 

The ejectile wave function, $\chi^{(-)}$, is written in terms of its positive 
energy component $\Psi_{f+}$ following the direct Pauli reduction method
\begin{eqnarray}
 \chi^{(-)}  =  \left(\begin{array}{c} 
{\displaystyle \chi_{+}} \\ 
\frac {\displaystyle 
\mbox{\boldmath $\sigma$}\cdot \mbox{\boldmath $p^{\, \prime}$}} 
{\displaystyle M+ E'+S-V} \ 
        \chi_{+} \end{array}\right) \ ,
\end{eqnarray}
where $S=S(r)$ and $V=V(r)$ are the scalar and vector potentials for the nucleon
with energy $E'$ \cite{cooper}. The upper component $\chi_{+}$ is related to 
a Schr\"odinger equivalent wave function $\Phi_{f}$ by the Darwin factor $D(r)$,
i.e.,
\begin{eqnarray}
\chi_{+} = \sqrt{D(r)}\Phi_{f} \ , \ \ \ \ \ \ 
D(r) = \frac{M+E'+S-V}{M+E'} \ .
\end{eqnarray}
$\Phi_{f}$ is a two-component wave function which is solution of a 
Schr\"odinger
equation containing equivalent central and spin-orbit potentials obtained from
the scalar and vector potentials. 

The choice of the electromagnetic operator is a longstanding problem. Here
we discuss the three current conserving expressions
\cite{defo,kelly}
\begin{eqnarray}
\widehat j_{\mathrm {cc1}}^{\mu} &=& G_M(Q^2) \gamma ^{\mu} - 
             \frac {\kappa}{2M} F_2(Q^2)\overline P^{\mu} \ , \nonumber \\
\widehat j_{\mathrm {cc2}}^{\mu} &=& F_1(Q^2) \gamma ^{\mu} + 
             i\frac {\kappa}{2M} F_2(Q^2)\sigma^{\mu\nu}q_{\nu} \ ,
	     \label{eq.cc} \nonumber \\
\widehat j_{\mathrm {cc3}}^{\mu} &=& F_1(Q^2) \frac{\overline P^{\mu}}{2M} + 
             \frac {i}{2M} G_M(Q^2)\sigma^{\mu\nu}q_{\nu} \ , 
\end{eqnarray}
where $q^{\mu} = (\omega,\q)$ is the four-momentum transfer,
$Q^2=\mid\q\mid^2-\omega ^2$, $\overline P^{\mu} = (E+E',\pm+\pf)$, 
$\kappa$ is the anomalous part of the magnetic
moment, $F_1$ and $F_2$ are the Dirac and Pauli nucleon form factors, $G_M =
F_1+\kappa F_2$ is the Sachs nucleon magnetic form factor, and
$\sigma^{\mu\nu}=\left(i/2\right)\left[\gamma^{\mu},\gamma^{\nu}\right]$. These 
expressions are equivalent for on-shell particles thanks to Gordon identity. 
However, since nucleons in the nucleus are off-shell we expect that these 
formulas should give different results. Current 
conservation is restored by replacing the longitudinal current and the bound 
nucleon energy by  \cite{defo}
\begin{eqnarray}
J^L = J^z = \frac{\omega}{\mid\q\mid}~J^0 \ , \ \ \ \ \ \ 
E = \sqrt{\mid \pm \mid^2 + M^2} = \sqrt{ \mid \pf-\q\mid^2 + M^2} \ .
\end{eqnarray}

\subsection{Nonrelativistic current} \label{sec.nr}

In nonrelativistic DWIA the transition amplitude of Eq. (\ref{eq.rj}) is 
evaluated 
using eigenfunctions of a Schr\"odinger equation
for both the bound and scattering states. 
In standard DWIA analyses phenomenological ingredients are usually adopted. In 
this work and in order to perform a 
consistent comparison with RDWIA calculations, we employ for the bound state 
the upper component of the Dirac wave function $\varphi$ and for the final state
the Schr\"odinger-like wave function $\Phi_{f}$. 

The nuclear current operator is obtained from the Foldy-Wouthuysen reduction
of the free-nucleon Dirac current through an expansion in a power series of 
$1/M$.
Performing the expansion through second order we get 
\begin{eqnarray}
j_{(0)}^0 &=& F_1 \ \ \ , \ \ j_{(1)}^0 = 0 \ \ \ , \ \ 
j_{(2)}^0 = \frac {\left(F_1+2\kappa F_2 \right) }{8M^2}\  \left(-Q^2 
  -  i \ss \cdot \P \times \q \right)\ ,  \nonumber \\
\j _{(0)} &=& 0 \ \ \ ,  \ \ 
\j _{(1)} = \frac {\left(F_1+\kappa F_2 \right)}{2M}\ i
\ss \times \q + \frac {1}{2M}\ F_1 \P \ \ \ , \ \ 
\j_{(2)} = -\frac {\left(F_1+2\kappa F_2 \right)}{8M^2}\  i \omega
 \ss \times \P \ . 
\end{eqnarray}

\subsection{The $\left(e,e^{\prime}p\right)$ cross section} \label{sec.eep2}

The coincidence cross section of the $\left(e,\ep p\right)$ reaction can be 
written 
as the contraction between the lepton tensor, completely determined by
quantum electrodynamics, and the hadron tensor, whose components are given by
suitable bilinear combinations of the nuclear transition amplitude in 
Eq. (\ref{eq.rj}). In case of unpolarized reactions the cross section can be
written in terms of four response functions,
$R_{\lambda\lambda^{\prime}}$, as
\inieq
\sigma  = 
K\ \left\{v _{L}
R_{L} +  v_{T}R_{T}+v_{LT}R_{LT}\cos\left(\vartheta\right) 
 +v_{TT}R_{TT}\cos\left(2\vartheta\right)\right\} \ ,  
\label{eq.fcs}
\fineq
where $K$ is a kinematic factor, and $\vartheta $ is the out-of-plane 
angle between the electron scattering plane and the $(\q, \pf)$ plane. The 
coefficients $v_{\lambda\lambda'}$ are obtained from the lepton tensor
components and depend only upon the electron kinematics \cite{book,kellyrep}.
The response functions are defined as
\inieq
R_{L} \propto \langle J^0 \left(J^0\right)^{\dagger }\rangle \ \ \ &,& \ \ 
R_{T} \propto \langle J^x \left(J^x\right)^{\dagger} \rangle +
          \langle J^y \left(J^y\right)^{\dagger} \rangle \ , \nonumber \\
R_{LT} \propto -2\ {\rm Re}
   \left[\langle J^x \left(J^0\right)^{\dagger }\rangle\right] \ \ \ &,& \ \ 
R_{TT} \propto  \langle J^x \left(J^x\right)^{\dagger} \rangle -
          \langle J^y \left(J^y\right)^{\dagger} \rangle \ , \label{eq.resf}
\fineq
where 
average over the initial and sum over
the final states is performed fulfilling energy conservation. 
In our frame of reference the $z$ axis is along $\q$, and the $y$ axis is
parallel to $\q\times\pf$.

If the electron beam is longitudinally polarized with helicity $h$, the
coincidence cross section for a knocked out nucleon with spin directed along
$\hat {\s}$ can be written as
\begin{eqnarray}
\sigma _{\text {h},\hat {\bf {\text s}}} =  \frac {1}{2}\ \sigma \ 
\bigg [1 + \P \cdot \hat {\s} + h 
\Big (A + \P ^{\prime }\cdot \hat {\s}\Big )\bigg ], \label{eq.polcs}
\end{eqnarray}
where $\sigma $ is the unpolarized cross section of Eq.~(\ref {eq.fcs}), $\P$ 
the induced polarization, $A$ the electron analyzing power and $\P ^{\prime }$
the polarization transfer coefficient. We choose for the polarimeter the three 
perpendicular directions: $\L$ parallel to $\pf$, $\N$ along 
$\q \times \pf $, and $\T = \N \times \L$.
\subsection{Results and Discussion}
\begin{figure}[th!]
\vskip -1.cm
\includegraphics[height=10cm, width=15cm]{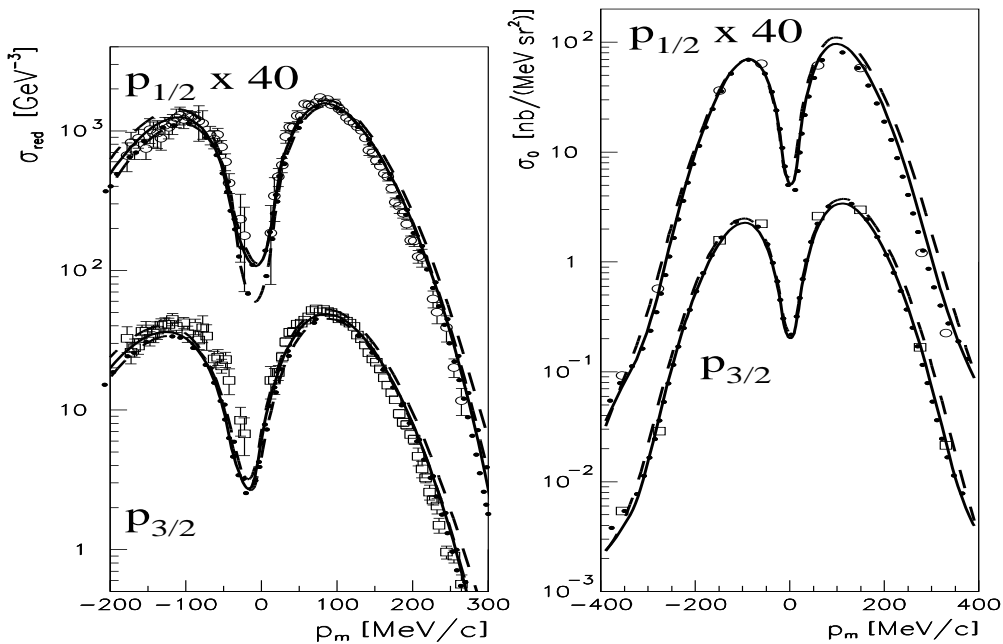} 
\caption{Left panel: reduced cross section for the $^{16}$O$(e,e'p)^{15}$N
transition to the $p\half$ ground state and $p\thalf$ first excited state of 
$^{15}$N at $E_p = 90$ MeV constant proton energy in the center-of-mass 
system in parallel kinematics~\protect\cite{nikhef}. Right panel: cross 
section for the same reaction 
but at $Q^2=0.8$ (GeV/$c)^2$ in constant $(\q, \omega)$ 
kinematics \cite{e89003}. Data 
for the $p\half$ state have been multiplied by 40.
Dashed, solid, and 
dotted lines represent the results of the RDWIA approach with cc1, cc2, cc3 
off-shell prescriptions, respectively. 
Dot-dashed lines in the left panel are the nonrelativistic results. } 

\label{fig:fig1}
\end{figure}

In Fig. \ref{fig:fig1} the unpolarized $^{16}$O$(e,e'p)$
reaction leading to the $p\half$ ground state and the 
$p\thalf$ first excited state of $^{15}$N is considered. 
In the left panel, data have been 
collected at NIKHEF in parallel kinematics at a constant proton
energy of 90 MeV in the center-of-mass system \cite{nikhef}.
They are presented in the form of the reduced cross section \cite{book}. 
The results for the 
transition to the $p\half$ ground state have been multiplied by 40.
The dot-dashed lines refer to the nonrelativistic calculations of 
Sec. \ref{sec.nr}. The theoretical results have been rescaled in order to
reproduce 
the data by applying the spectroscopic factors 
$Z_{p\half}=0.64$ and $Z_{p\thalf}=0.54$, respectively.
The solid lines show the results of the RDWIA analysis 
with the cc2 off-shell prescription; dashed and
dotted lines indicate the results when using the cc1 and cc3 recipes,
respectively. 
The resulting spectroscopic factors, $Z_{p\half}= 0.708$ and 
$Z_{p\thalf}=0.602$, have been obtained by a $\chi^2$ fit using the cc3 
current \cite{mrd}, which gives an overall better description of the $(e,e'p)$ 
observables. Only small differences are found between the relativistic and the
nonrelativistic models. Thus, they
are almost equivalent in comparison with the data, which are reasonably
described by both calculations. 

In the right panel, the same reaction is considered
at the JLab constant $(\q, \omega)$ kinematics
with $Q^2 = 0.8$ (GeV/$c$)$^2$~\cite{e89003}. 
The data now refer to the differential unpolarized cross section. The 
$p\half$ results are multiplied 
by a factor 40. The theoretical curves have the same meaning as in the left 
panel and are rescaled again by the same spectroscopic factors. Only RDWIA
calculations are shown since at the proton energy of this experiment
relativistic effects are large and a nonrelativistic analysis gives unreliable
results.
The agreement 
with the data is very good also in this case. This outcome is particularly 
welcome, since the spectroscopic factors correspond to a nuclear property that 
must be independent of $Q^2$. 

\begin{figure}[t!]
\vskip -1.cm
\includegraphics[height=11cm, width=15cm]{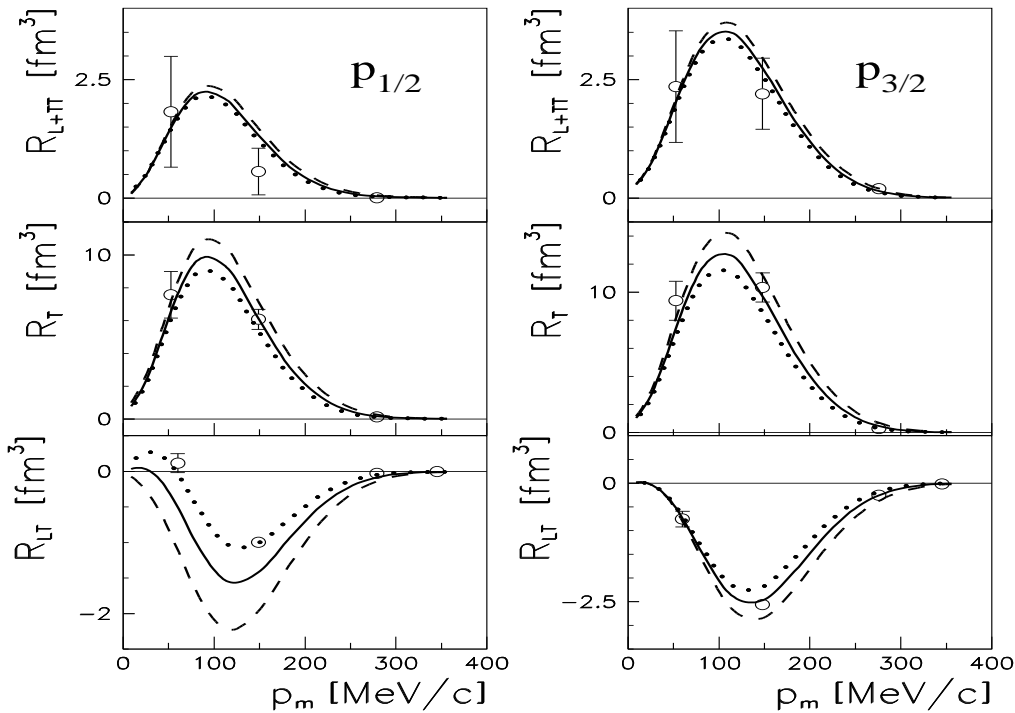} 
\caption{Response functions for the $^{16}$O$(e,e'p)$ reaction at 
$Q^2=0.8$ (GeV/$c)^2$ in constant $(\q, \omega)$  kinematics \cite{e89033}
leading to the $^{15}$N $p\half$ (left column) and $p\thalf$ (right column) 
residual states.   
Line convention as in Fig. \ref{fig:fig1} } 
\label{fig:fig2}
\end{figure}

\begin{figure}[t!]
\vskip -0.5cm
\begin{center}
\includegraphics[height=11cm, width=15cm]{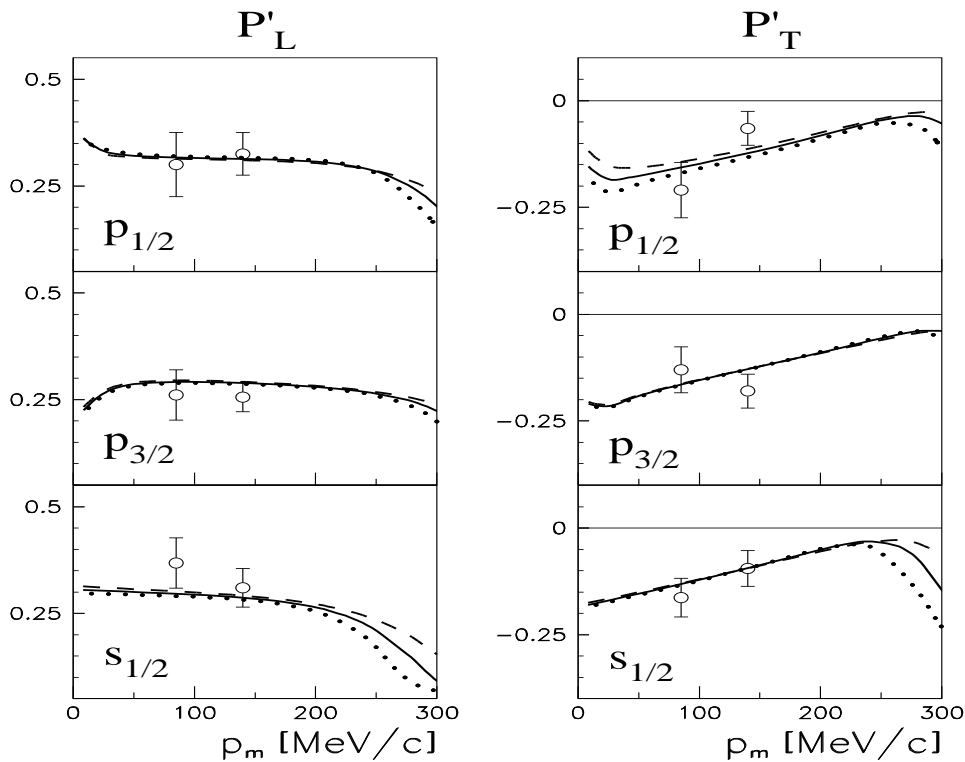} 
\caption {Polarization transfer components $P^{\prime \, L}, P^{\prime \, T}$ 
for the $^{16}$O($\vec e,e'\vec p$) reaction at 
$Q^2=0.8$ (GeV/$c)^2$ in constant $(\q, \omega)$ kinematics \cite{e89033}
leading to the 
$^{15}$N $p\half$, $p\thalf$ and $s\half$ residual states. 
Line convention as in Fig. \ref{fig:fig1}.}
\label{fig:fig3}
\end{center}
\end{figure}

In Fig. \ref{fig:fig2} the response functions measured at JLab in the same
kinematics with $Q^2 = 0.8$ (GeV/$c$)$^2$~\cite{e89003} are displayed and
compared with our RDWIA calculations. The agreement with the data is
satisfactory and of about the same quality as in other relativistic analyses
\cite{spain,kelly},
but for the $R_{LT}$ response function, where only the cc3 calculation 
reproduces
the $p\half$ data at low missing momentum while the cc2 one better reproduces 
the $p\thalf$ data.
In Fig. \ref{fig:fig3} the polarization transfer
components $P^{\prime \, L}, P^{\prime \, T}$  are shown as 
functions of the missing momentum $p_m$ for 
the $^{16}$O($\vec e,e'\vec p$) reaction at $Q^2=0.8$ (GeV/$c)^2$ and
constant $(\q, \omega)$ for the transitions to the $^{15}$N 
$p\half$, $p\thalf$, and $s\half$, respectively \cite{e89033}. 
For these observables and in this
kinematics, the sensitivity to off-shell effects is at most $\lesssim 15$\%.  
The overall agreement with the data is still good.

As a first step to study the role of meson exchange currents (MEC),
we have considered in Ref. \cite{mgpseag}  
the contribution due to the seagull diagram.
The seagull current is written in momentum space as in 
Refs. \cite{tjon,tjon2} and with
the cutoff $\Lambda = 1250$ MeV in the pion propagator.
The inclusion of the seagull diagram enhances the RDWIA results,
but, in contrast to $\left(\gamma,p\right)$ scattering, the effects 
are generally small and visible only at high missing momenta. Thus, the
comparison with data, that were already well reproduced by the direct knoclout
 model, is
practically unaffected. In particular, no
significant effects were obtained on the polarization observables from 
MIT-Bates
on $^{12}$C$( e, e^{\, \prime} \vec p\ \!)^{11}$B \cite{batespn} and from JLab 
on $^{16}$O$(\vec e, e^{\, \prime} \vec p\ \!)^{15}$N \cite{e89033} 
reactions.

\section{The $\left(e,e^{\prime}\right)$ reaction} \label{sec.ee}

In the one photon exchange approximation the inclusive cross section for the
quasielastic $(e,e^{\prime})$ scattering on a nucleus is given in terms of two
response function
as \cite{book}
\inieq
\sigma_{{{inc}}} = K \left( 2\varepsilon_{{L}}R_{{L}}
 + R_{{T}}\right) \ , \label{eq.cs}
\fineq 
where $K$ is a kinematical factor and 
\inieq
\varepsilon_{{L}} = \frac{Q^2}{\q^2} \left( 1 + 2 \frac{\q^2}{Q^2}
\tan^2{(\vartheta_e/2)}\right)^{-1} \ \label{eq.polar}
\fineq
measures the polarization of the virtual photon. In Eq. (\ref{eq.polar})
$\vartheta_e$ is the scattering angle of the electron. 
The longitudinal and transverse response
functions, $R_{{L}}$ and $R_{{T}}$, contain all nuclear structure information
and are defined as in Eq. (\ref{eq.resf}). 
They are directly related to the hadron tensor components
\inieq
W^{\mu\mu}(\omega,q) = 
\! \! \bint\sum_{ {f}} \!  \mid\! \langle 
\Psi_{{f}}\mid J_N^{\mu}(\q) \mid \Psi_0\rangle\! \mid^{2}
\delta (E_0 +\omega - E_{ {f}})  
=-\frac{1}{\pi} \textrm{Im} \langle \Psi_0
\mid \! J_N^{\mu\dagger}(\q) G(E_{ {f}}) J_N^{\mu}(\q) \! 
\mid \Psi_0 \rangle  . 
\label{eq.hadrontensor}
\fineq
Here $J_N^{\mu}$ is the nuclear charge-current operator which connects the 
initial
state $\mid\Psi_0\rangle$ of the nucleus, of energy $E_0$, with the final states
$\mid\Psi_{ {f}} \rangle$, of energy $E_{ {f}}$, both
eigenstates of the $(A+1)$-body Hamiltonian $H$. The sum runs over the
scattering states corresponding to all of the allowed asymptotic configurations
and includes possible discrete states.  
$G(E_{{f}})$ is the Green
function related to $H$, i.e.,
\inieq
G(E_{ {f}}) = \frac{1}{E_{ {f}} - H + i\eta} \ . \label{eq.green}
\fineq
Here and in all the equations involving $G$ the limit for $\eta \rightarrow
+0$ is understood. It must be performed after calculating the matrix elements 
between normalizable states.

In Ref. \cite{ee} we have showed that the spectral representation of the single
particle Green's functions related to the optical potential allows practical 
calculations of the hadron tensor components for the inclusive $(e,e')$ 
scattering.
Here, we briefly recall the most important points of the method, without
discussing in details the approximations involved. Assuming only one-body terms
in the nuclear current, Eq. (\ref{eq.hadrontensor}) can be reduced to a 
single-particle expression whose self-energy is the Feshbach's optical 
potential \cite{ee}. We then perform a biorthogonal expansion of the full
particle-hole Green's function 
in terms of the eigenfunctions the optical potential, $\mcv$, and its 
Hermitian conjugate, $\mcv^{\dagger}$, i.e.,
\inieq 
[\mathcal{E} - T -\mcv^{\dagger}(E)] \mid
\chi_{\mathcal{E}}^{(-)}(E)\rangle = 0 \ \ \ ,\ \ 
\left[\mathcal{E} - T - \mcv(E)\right] \mid
\tilde {\chi}_{\mathcal{E}}^{(-)}(E)\rangle = 0 \ . \label{eq.inco2}
\fineq
Note that ${\mathcal{E}}$ and $E$ are not necessarily the same.
The spectral representation is
\inieq
\mcg(E) = \int_M^{\infty} \diff \mathcal{E}\mid\tilde
{\chi}_{\mathcal{E}}^{(-)}(E)\rangle 
\frac{1}{E-\mathcal{E}+i\eta} \langle\chi_{\mathcal{E}}^{(-)}(E)\mid 
\ , \label{eq.sperep}
\fineq
and the hadron tensor can be written in an expanded version in terms of the
single-partcile wave function, $\mid\varphi_n\rangle$, of the initial 
state, corresponding to the energy $\varepsilon_n$ and whose spectral 
strength is $\lambda_n$, as
\inieq
W^{\mu\mu}(\omega , q) = -\frac{1}{\pi} \sum_n  \textrm{Im} \bigg[
 \int_M^{\infty} \diff \mathcal{E} \frac{1}{E_{
{f}}-\varepsilon_n-\mathcal{E}+i\eta}  
  T_n^{\mu\mu}(\mathcal{E} ,E_{{f}}-\varepsilon_n) \bigg]
\ , \label{eq.pracw}
\fineq
where
\inieq
T_n^{\mu\mu}(\mathcal{E} ,E) = \lambda_n\langle \varphi_n
\mid j^{\mu\dagger}(\q) \sqrt{1-\mcv'(E)}
\mid\tilde{\chi}_{\mathcal{E}}^{(-)}(E)\rangle 
 \langle\chi_{\mathcal{E}}^{(-)}(E)\mid  \sqrt{1-\mcv'(E)} j^{\mu}
(\q)\mid \varphi_n \rangle  \ . \label{eq.tprac}
\fineq
In Ref. \cite{ee} we have shown that the factor $\sqrt{1-\mcv'(E)}$ accounts for
interference effects between different channels and allows the replacement of
the mean field $\mcv$ with the phenomenological optical potential 
$\mcv_{ L}$. In Eq. (\ref{eq.tprac}) $\mcv'(E)$ is the energy derivative of the 
mean field potential.
After calculating the limit for $\eta \rightarrow +0$ 
Eq. (\ref{eq.tprac}) reads
\inieq
W^{\mu\mu}(\omega , q) = \sum_n \Bigg[ \textrm{Re} T_n^{\mu\mu}
(E_{{f}}-\varepsilon_n, E_{{f}}-\varepsilon_n)  
- \frac{1}{\pi} \mathcal{P}  \int_M^{\infty} \diff \mathcal{E} 
\frac{1}{E_{{f}}-\varepsilon_n-\mathcal{E}} 
\textrm{Im} T_n^{\mu\mu}
(\mathcal{E},E_{{f}}-\varepsilon_n) \Bigg] \ , \label{eq.finale}
\fineq
where $\mathcal{P}$ denotes the principal value of the integral. 
Eq. (\ref{eq.finale}) separately involves the real and imaginary parts of 
$T_n^{\mu\mu}$.   

Let us examine the expression of $T_n^{\mu\mu}(\mathcal{E},E)$ at 
$\mathcal{E}=E=E_{\textrm{f}}-\varepsilon_n$ for a fixed $n$. This is the most 
important case since it appears in the first term in the right hand side of 
Eq. (\ref{eq.finale}), which gives the main contribution. 
Disregarding the square root correction, due to interference effects, one
observes that in Eq. (\ref{eq.tprac}) the second matrix element (with the
inclusion of $\sqrt{\lambda_n}$) is the transition amplitude for the single
nucleon knockout from a nucleus in the state $\mid \Psi_0\rangle$ leaving the
residual nucleus in the state $\mid n \rangle$. The attenuation of its strength,
mathematically due to the imaginary part of $\mcv^{\dagger}$, is related to the
flux lost towards the channels different from $n$. In the inclusive response
this attenuation must be compensated by a corresponding gain due to the flux
lost, towards the channel $n$, by the other final states asymptotically
originated by the channels different from $n$. In the description provided by
the spectral representation of Eq. (\ref{eq.finale}), the compensation is
performed by the first matrix element in the right hand side of 
Eq. (\ref{eq.tprac}), where the imaginary part of $\mcv$ has the effect of 
increasing
the strength. Similar considerations can be made, on the purely mathematical
ground, for the integral of Eq. (\ref{eq.finale}), where the 
amplitudes involved in $T_n^{\mu\mu}$ have no evident physical meaning as 
$\mathcal{E}\neq E_{\textrm{f}}-\varepsilon_n$. 
We want to stress
here that in the Green function approach it is just the imaginary part of 
$\mcv$ which accounts for the redistribution of the strength among different 
channels.

The cross sections and the response functions of the inclusive quasielastic
electron scattering are calculated from the single particle expression of the
hadron tensor in  Eq. (\ref{eq.finale}). After the replacement of the 
mean field $\mcv(E)$ by the empirical optical model potential 
$\mcv_{ {L}}(E)$, the matrix elements of the nuclear current operator in 
Eq. (\ref{eq.tprac}), which represent the main ingredients of the calculation, 
are of the same kind as those giving the RDWIA transition amplitudes of the 
$(e,e'p)$ reaction. Thus, the same treatment can be used to describe 
the initial and final state wave functions and the one-nucleon electromagnetic
current.

\subsection{Results and discussion} 

\begin{figure}[t]
\vskip -1.15cm
\includegraphics[height=10cm, width=17cm]{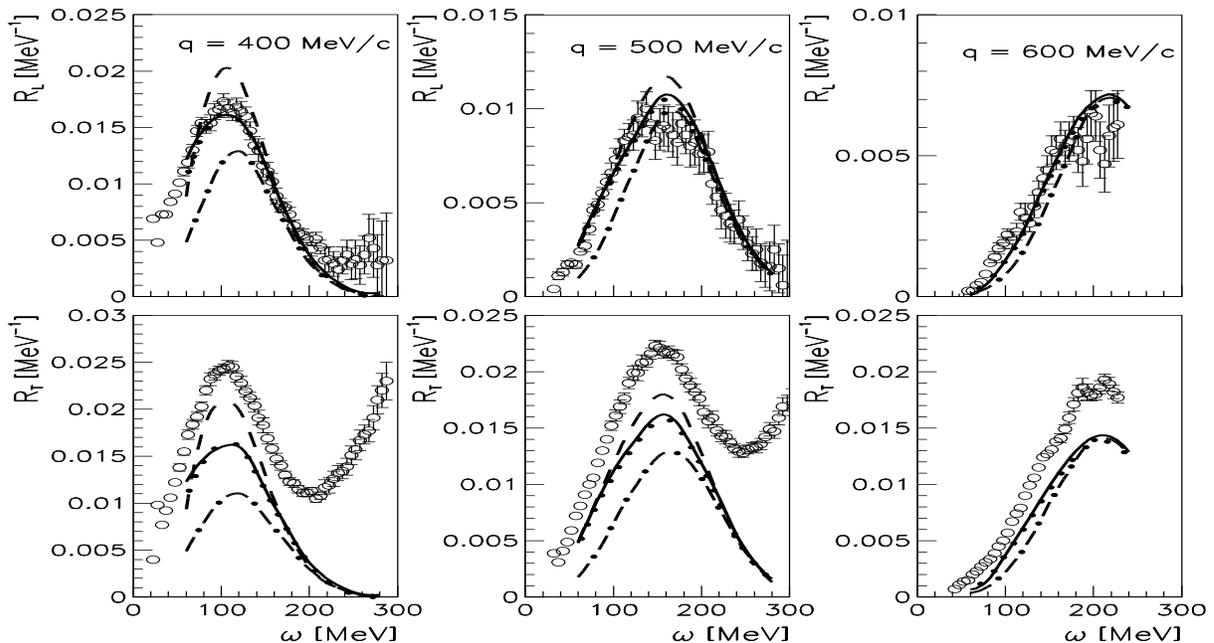} 
\vskip -0.5cm
\caption {Longitudinal (upper panel) and transverse (lower panel) response 
functions for the
$^{12}$C$(e,e')$ reaction at $q = 400$, $500$, and $600$ MeV$/c$, respectively.  
Solid and dotted lines represent the relativistic results with and without the 
inclusion 
of the factor in Eq. (\ref{eq.def}), respectively. Dashed lines give the result 
without the integral in Eq. (\ref{eq.finale}).
Dot-dashed lines are the contribution of integrated single nucleon knockout
only. The data are from Ref. \cite{saclay}.}
\label{fig1}
\end{figure}

\begin{figure}[t]
\vskip -1.cm
\includegraphics[height=15cm, width=15cm]{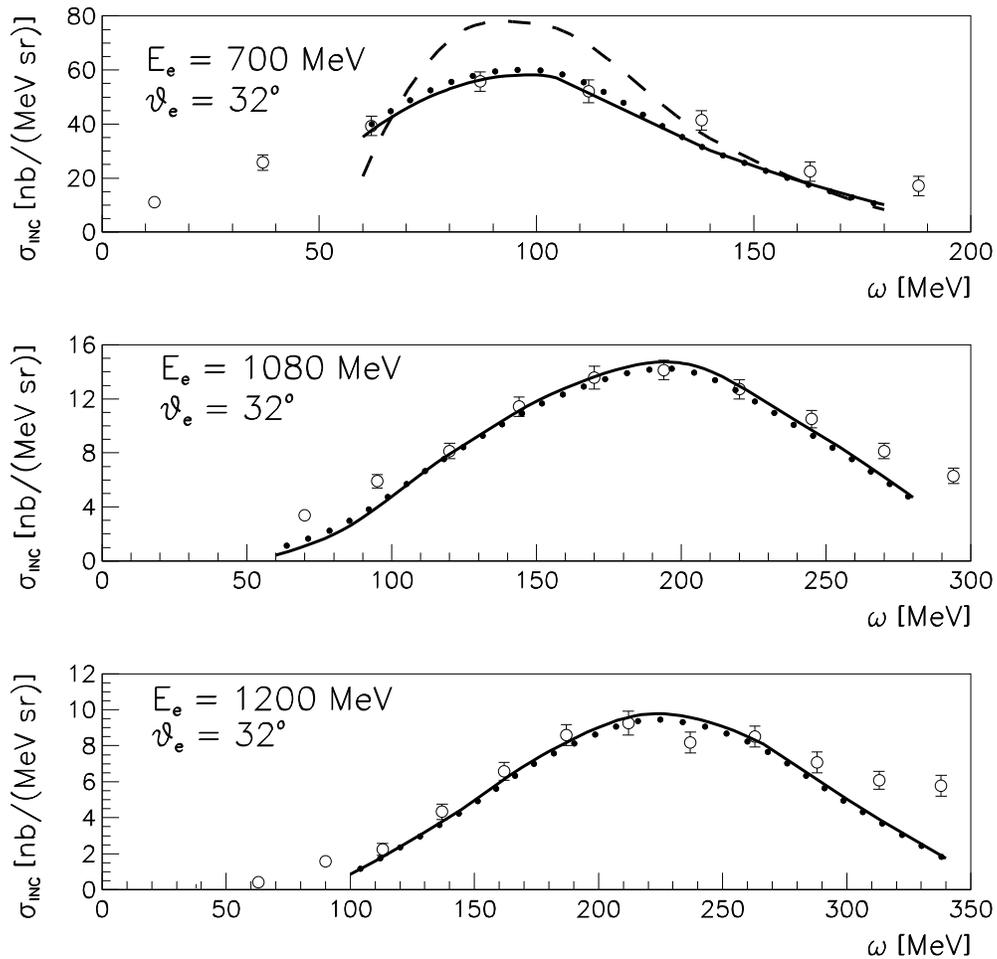} 
\vskip -0.5cm
\caption {The cross section for the inclusive $^{16}$O$(e,e')$ reaction at 
$\vartheta_e = 32^{\mathrm o}$ and $E_e = 700$, $1080$, and $1200$ MeV. 
The data are from ADONE-Frascati \cite{csfrascati}. Line convention as in 
Fig. \ref{fig1}.}
\label{fig6}
\end{figure}


The longitudinal and transverse response functions for $^{12}$C at 
$q = 400$ MeV$/c$ are displayed in Fig. \ref{fig1} (left column) and compared 
with the Saclay 
data \cite{saclay}. The low energy 
transfer values are not given because the relativistic optical 
potentials are not available at low energies.

The agreement with the data is generally satisfactory for the longitudinal 
response. The transverse response is underestimated. This
is a systematic result of the calculations and was also found in the
nonrelativistic approach of Ref. \cite{capuzzi}. It may be attributed to
physical effects which are not considered in the single-particle Green 
function approach, e.g., meson exchange currents. 

The effect of the integral in Eq. (\ref{eq.finale}) is also displayed.
Its contribution is 
important and essential to reproduce the experimental longitudinal response.

The contribution arising from interference between different channels gives 
rise to the factor 
\inieq
 \sqrt{1- \mcv'_{{L}}(E)}= \sqrt{1- \beta S'(E) - V'(E)} \ .
 \label{eq.def}
\fineq
We see, however, that here it gives only a slight contribution, due to a 
compensation between the energy derivatives $S'(E)$ and $V'(E)$. 

The contribution from all the integrated single nucleon knockout
channels is also drawn in Fig. \ref{fig1}. It is significantly smaller than the 
complete calculation. The reduction, which is larger at lower values of
$\omega$, gives an indication of the relevance of inelastic channels.  

The longitudinal and transverse response functions for $^{12}$C at $q = 500$
(middle column)
and $q = 600$ MeV$/c$ (right column) are also displayed and compared 
with the Saclay data \cite{saclay} in Fig. \ref{fig1}. 
As already found at $q = 400$ MeV$/c$, a good agreement with 
the data is obtained in both cases for the longitudinal response, while the 
transverse response is underestimated. 
Only a slight effect is given by the 
factor in Eq. (\ref{eq.def}) arising from the interference between different
channels. The role of the integral in Eq. (\ref{eq.finale}) decreases 
increasing 
the momentum transfer. 

In Fig. \ref{fig6} we consider the $^{16}$O$(e,e')$ inclusive cross section data 
taken at ADONE-Frascati \cite{csfrascati} with beam energy ranging from $700$ 
to $1200$ MeV and a scattering angle $\vartheta_e \simeq 32^{\mathrm o}$. The 
NLSH wave 
functions have been used in the calculations. The agreement with data is 
good in all the situations considered. The integral in Eq. (\ref{eq.finale}) 
produces a 
reduction which is now essential to reproduce the data at $700$ MeV, which 
correspond to a momentum transfer  $q \lesssim 400$ MeV$/c$. Its
contribution can be neglected when $q \simeq 600$ 
MeV$/c$. The effect of the factor in Eq. (\ref{eq.def}) is very small.



\begin{thebibliography}{}



\bibitem{frou}
S. Frullani and J. Mougey, 
Adv. Nucl. Phys. {\bf 14}, 1 (1984).

\bibitem{mou}
J. Mougey {\it et al.}, 
Nucl. Phys. A {\bf 262}, 461 (1976).

\bibitem{bern}
M. Bernheim {\it et al.}, 
Nucl. Phys. A {\bf 375}, 381 (1982).

\bibitem{dewitt}
P.K.A. de Witt Huberts, 
J. Phys. G {\bf 16}, 507 (1990).

\bibitem{lapik}
L. Lapik\'as, 
Nucl. Phys. A {\bf 553}, 297c (1993).

\bibitem{blomq}
K.I. Blomqvist {\it et al.},
Phys. Lett. B {\bf 344}, 85 (1995).
 
 \bibitem{nikhef}
M. Leuschner {\it et al.}, 
Phys. Rev. C {\bf 49}, 955 (1994).

\bibitem{DWEEPY}
C. Giusti and F.D. Pacati, 
Nucl. Phys. A {\bf 473}, 717 (1987); A {\bf 485}, 461 
(1988).

\bibitem{book}
S. Boffi, C. Giusti, F.D. Pacati, and M. Radici,
{\it Electromagnetic Response of Atomic Nuclei}, Oxford Studies in Nuclear
Physics Vol. 20 (Clarendon Press, Oxford, 1996).


\bibitem{kellyrep}
J.J. Kelly,
Adv. Nucl. Phys. {\bf 23}, 75 (1996).


\bibitem{pickvanord}
A. Picklesimer and J.W. Van Orden,
Phys. Rev. C {\bf 40}, 290 (1989).

\bibitem{jinonl}
Y. Jin and D.S. Onley,
Phys. Rev. C {\bf 50}, 377 (1994).

\bibitem{spain}
J.M. Ud\'{\i}as, J.A. Caballero, E. Moya de Guerra, J.R. Vignote, and A. Escuderos, 
Phys. Rev. C {\bf 64}, 024614 (2001).

\bibitem{heda}
M. Hedayati-Poor, J.I. Johansson, and H.S. Sherif, 
Phys. Rev. C {\bf 51}, 2044 (1995).

\bibitem{kelly}
J.J. Kelly,
Phys. Rev. C {\bf 60}, 044609 (1999).

\bibitem{mgp}
A. Meucci, C. Giusti, and F.D. Pacati,
Phys. Rev. C {\bf 64}, 014604 (2001).

\bibitem{e89003}
J. Gao {\it et al.}, 
Phys. Rev. Lett. {\bf 84}, 3265 (2000).

\bibitem{e89033}
S. Malov {\it et al.},
Phys. Rev. C {\bf 62}, 057302 (2000).

\bibitem{bgpc}
S. Boffi, C. Giusti, F.D. Pacati, and F. Cannata, 
Nuovo Cimento A {\bf 98}, 291 (1987).

\bibitem{meu}
A. Meucci,  
Phys. Rev. C {\bf 65}, 044601 (2002).

\bibitem{mrd}
M. Radici, A. Meucci, and W.H. Dickhoff,
Eur. Phys. J. A {\bf 17}, 65 (2003). 

\bibitem{ee}
A. Meucci, F. Capuzzi, C. Giusti, and F.D. Pacati,
Phys. Rev. C {\bf 67}, 054601 (2003).

\bibitem{batesca}
C.F. Williamson {\it et al.},
Phys. Rev. C {\bf 56}, 3152 (1997).

\bibitem{csfrascati}
M. Anghinolfi {\it et al.},
Nucl. Phys. A {\bf 602}, 405 (1996).

\bibitem{jlabpro}
J.P. Chen, S. Choi, and Z.E. Meziani, spokespersons, JLab experiment E-01-016.

\bibitem{Sluys}
V. Van der Sluys, J. Ryckebusch, and M. Waroquier,
Phys. Rev. C {\bf 51}, 2664 (1995).

\bibitem{Cenni}
R. Cenni, F. Conte, and P. Saracco,
Nucl. Phys. A {\bf 623}, 391 (1997).

\bibitem{Amaro}
J.E. Amaro, M.B. Barbaro, J.A. Caballero, T.W. Donnelly, and A. Molinari,
Phys. Rep. {\bf 368}, 317 (2002); Nucl. Phys. A {\bf 723}, 181 (2003).

\bibitem{Fabrocini}
A. Fabrocini,
Phys. Rev. C {\bf 55}, 338 (1997).

\bibitem{Co}
G. Co' and A.M. Lallena, Ann. Phys. {\bf 287}, 101 (2001).

\bibitem{Leidemann}
W. Leidemann and G. Orlandini,
Nucl. Phys. A {\bf 506}, 447 (1990).

\bibitem{Sick}
I. Sick, in {\it Nuclear Theory - Proceedings of the XXI International 
Workshop on Nuclear Theory}, edited by V. Nikolaev (Heron Press Science 
Series, Sofia, 2002), p. 16.

\bibitem{lala}
G.A. Lalazissis, J. K\"onig, and P. Ring, 
Phys. Rev. C {\bf 55}, 540 (1997).

\bibitem{cooper}
E.D. Cooper, S. Hama, B.C. Clark, and R.L. Mercer, 
Phys. Rev. C {\bf  47}, 297 (1993).

\bibitem{defo}
T. de Forest, Jr., 
Nucl. Phys. A {\bf 392}, 232 (1983).



\bibitem{mgpseag}
A. Meucci, C. Giusti, and F. D. Pacati, 
Phys. Rev. C {\bf 66}, 034610 (2002).


\bibitem{tjon}
M.J. Dekker, P.J. Brussaard, and J.A. Tjon, 
Phys. Rev. C {\bf 49}, 2650 (1994). 

\bibitem{tjon2}
G.H. Martinus, O. Scholten, and J.A. Tjon, 
Phys. Rev. C  {\bf 58}, 686 (1998).


\bibitem{batespn}
R.J. Woo {\it et al.},  
Phys. Rev. Lett. {\bf 80}, 456 (1998).

\bibitem{saclay}
P. Barreau {\it et al.},
Nucl. Phys. A {\bf 402}, 515 (1983); Note CEA N-2334.

\bibitem{capuzzi}
F. Capuzzi, C. Giusti, and F.D. Pacati,
Nucl. Phys. A {\bf 524}, 681 (1991).


\end{thebibliography}
\end{document}